\documentclass[aip,graphicx,rsi, amsmath,amssymb,reprint,floatfix]{revtex4-1}

\usepackage{graphicx}
\usepackage{dcolumn}
\usepackage{bm}
\usepackage[utf8]{inputenc}
\usepackage[T1]{fontenc}
\usepackage{mathptmx}
\usepackage{fixltx2e}
\usepackage{textcomp}
\usepackage{gensymb}

\usepackage{etoolbox}

\makeatletter
\appto\abstract{%
  \let\latexlist\list \rightskip=\leftskip
  \def\list{\edef\keeprightskip{\the\rightskip}\latexlist}%
  \patchcmd\latexlist{\ignorespaces}{\rightskip\keeprightskip\ignorespaces}{}{}%
}
\makeatother

\DeclareRobustCommand{\bigO}{%
  \text{\usefont{OMS}{cmsy}{m}{n}O}%
}

\draft 

\begin{document}


\title[High-Bandwidth, Variable-Resistance Differential Noise Thermometry]{High-Bandwidth, Variable-Resistance Differential Noise Thermometry}



\author{A. V. Talanov}
\affiliation{Department of Physics, Harvard University, Cambridge, Massachusetts 02138, USA}
\affiliation{John A. Paulson School of Engineering and Applied Sciences, Harvard University, Cambridge, MA 02138, USA}

\author{J. Waissman}
\affiliation{Department of Physics, Harvard University, Cambridge, Massachusetts 02138, USA}

\author{T. Taniguchi}
\affiliation{International Center for Materials Nanoarchitectonics, National Institute for Materials Science,  1-1 Namiki, Tsukuba 305-0044, Japan}

\author{K. Watanabe}
\affiliation{Research Center for Functional Materials, National Institute for Materials Science, 1-1 Namiki, Tsukuba 305-0044, Japan}

\author{P. Kim}
\homepage[]{http://kim.physics.harvard.edu}
\affiliation{Department of Physics, Harvard University, Cambridge, Massachusetts 02138, USA}
\affiliation{John A. Paulson School of Engineering and Applied Sciences, Harvard University, Cambridge, MA 02138, USA}


\date{\today}

\begin{abstract}
We develop Johnson noise thermometry applicable to mesoscopic devices with variable source impedance with high bandwidth for fast data acquisition. By implementing differential noise measurement and two-stage impedance matching, we demonstrate noise measurement in the frequency range 120-250~MHz with a wide sample resistance range 30~$\Omega$-$100~\text{k}\Omega$ tuned by gate voltages and temperature. We employ high-frequency, single-ended low noise amplifiers maintained at a constant cryogenic temperature in order to maintain the desired low noise temperature. We achieve thermometer calibration with temperature precision up to 650~$\mu$K on a 10~K background with 30~s of averaging. Using this differential noise thermometry technique, we measure thermal conductivity on a bilayer graphene sample spanning the metallic and semiconducting regimes in a wide resistance range, and we compare it to the electrical conductivity.
\end{abstract}


\maketitle 

\section{\label{sec:Introduction}Introduction}
Johnson noise thermometry (JNT) \cite{White1996} has broad applications ranging from fundamental science \cite{Qu2019}, to use in harsh environments such as nuclear reactors~\cite{Brixy1971Nuclear,Pearce2015Nuclear,Kisner2004}, and commerce/industry\cite{Bramley2016,Bramley2020,Qu2019}. JNT can function as a nanoscale probe to study thermal behavior in nanoscale devices and potentially help engineer more efficient cooling pathways to solve the growing problem of excess heat management in continuing miniaturization of electronics\cite{Pop2006,shabany2010heat}. Fundamentally, JNT allows measurement of the cosmic microwave background \cite{Jarosik2003,Smoot1992,Bersanelli2002,Kogut2004}, high-accuracy verification of the Boltzmann constant\cite{FlowersJacobs2017,Qu2017}, and extension of the International Temperature Scale (ITS) down to  6~mK~\cite{Soulen1994,Schuster2003,Fellmuth2003}.

The advantages of JNT include (1) functionality as a primary thermometer, (2) dependence only on the sample resistance and not on any other microscopic or macroscopic system parameters, (3) a wide temperature range of application, and (4) the ability to non-invasively probe nanoscale conductors by directly measuring their fluctuations, without perturbing the system as would resistive or other types of thermometry.

The Johnson noise of a resistor $R$ at temperature $T$ is the mean square of voltage fluctuations $v_N$ measured over some rectangular frequency band of width $\Delta f$ given by the formula $\langle v_N^2 \rangle=4k_B TR\Delta f$, where $k_B$ is the Boltzmann constant and the bracket indicates thermal averaging at an equilibrium state. The ideal uncertainty of a temperature measured via noise is given by the total-power Dicke radiometer formula \cite{Dicke1946,Wait1967,White1996,Qu2019}
\begin{align}
\sigma_T=\frac{T_\text{samp}+T_N}{\sqrt{\tau \Delta f_\text{c}}}.
\label{eq1}
\end{align}
where $T_\text{samp}$ is the sample temperature, $T_N$ is the effective amplifier (background) noise temperature, $\tau$ is the averaging time, and $\Delta f_\text{c}$ is the correlation bandwidth, related to $\Delta f$ and explained further below. In measuring high-temperature ($>100~\text{K}$) samples, amplifier noise is relatively small and not a limiting factor, but for colder samples ($<10~\text{K}$) amplifier noise can dominate the sample noise and necessitate long averaging times to achieve high precision with low-bandwidth thermometers\cite{Qu2019}.  Measuring the fluctuations at higher frequencies with larger bandwidth can overcome this limitation.

At radio frequencies (RF), electrical components such as cables and amplifiers are typically engineered to have an impedance of $Z_0=50~\Omega$, yet many mesoscopic samples possess higher resistance of order $R\sim h/e^2 =25~\text{k}\Omega$. Connecting such a sample directly to an amplifier mismatches the sample impedance ($R\not\sim Z_0$), 
where a fraction $\left| \Gamma \right| ^2=\left|\left(Z_0-Z\right)/\left(Z_0+Z\right)\right|^2$ of the noise power generated by the sample is reflected at the amplifier, and only the remaining fraction $1-\left|\Gamma\right|^2$ of the incident power is amplified, creating difficulty in measuring noise from high-resistance samples.

A single-stage inductor-capacitor ($L,C$) impedance matching circuit (MC) resolves this problem by transforming the bare sample resistance $R$ to a complex impedance $Z(R,f)$ that approximates the input impedance $Z_0$ of the amplifier over some frequency band, far away from which the noise coupling becomes inefficient~\cite{pozar2012microwave}. In mesoscopic samples, varying a field-effect gate, magnetic field, and other device parameters can vary the resistance over several orders of magnitude; therefore, in employing our JNT technique, we are chiefly interested in impedance matching over as large of a resistance range as possible.

\begin{figure}
\includegraphics{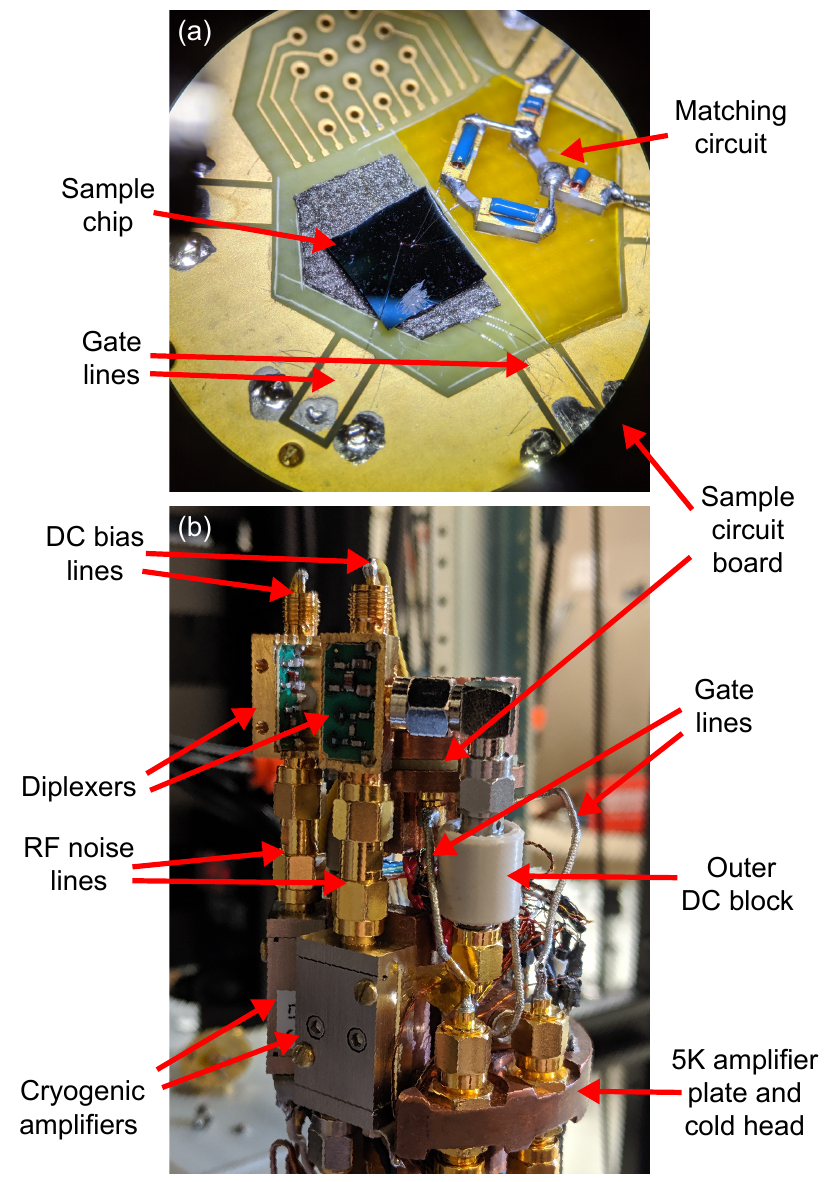}
\caption{\label{fig:Setup Picture} (a) Microscope image of the circuit board with the sample chip and matching circuit. (b) The cryocooler cold head with LNA and other low temperature measurement components. The sample enclosure is weakly thermally anchored to the 5~K plate with an aluminum tube; it is heated separately from the 5~K plate.}
\end{figure}

In this paper, we describe an approach to JNT for mesoscopic devices with highly variable resistance. Our differential measurement setup outperforms a single-ended setup, enabling symmetric biasing to remove unwanted thermoelectric currents. The two-stage lumped-element MC allows us to measure noise efficiently over several orders of magnitude of sample resistance. We describe the thermometer calibration procedure for variable-resistance samples. To demonstrate our technique, we measure the thermal conductance of a bilayer graphene (BLG) sample, at a temperature of 10~K, spanning four orders of magnitude of device resistance as we vary the gate voltages. The resultant uncertainty is consistent with the Dicke formula.

\section{\label{sec:level1}Noise Thermometer Circuit Design}
\subsection{\label{sec:level2}General Design Principles}
In designing our noise measurement circuit, our main goals are to measure noise over a wide range of sample resistance, while still maintaining a low background noise temperature $T_N$ and a wide bandwidth to facilitate faster measurements. We use commercially available components rather than custom ones wherever possible to simplify the design and to allow interchangeability and modifications. The MC described below in Section~\ref{Multi-Stage Matching} transforms a wide range of device impedance to closely match the 50~$\Omega$ RF components.  The circuit also allows separate quasi-DC (AC) and RF measurements on the device. The total noise power $P$ that we measure is the band-integrated quantity of the noise power spectral density $S$, with the high- and low-pass filters at $f_{1,2}$ determining the integration bounds:
\begin{align}
P=\int_{f_1}^{f_2}S(f)df.
\label{power_Integral_eq}
\end{align}

\subsection{\label{sec:level1}Circuit Description}
\begin{figure*}
\includegraphics{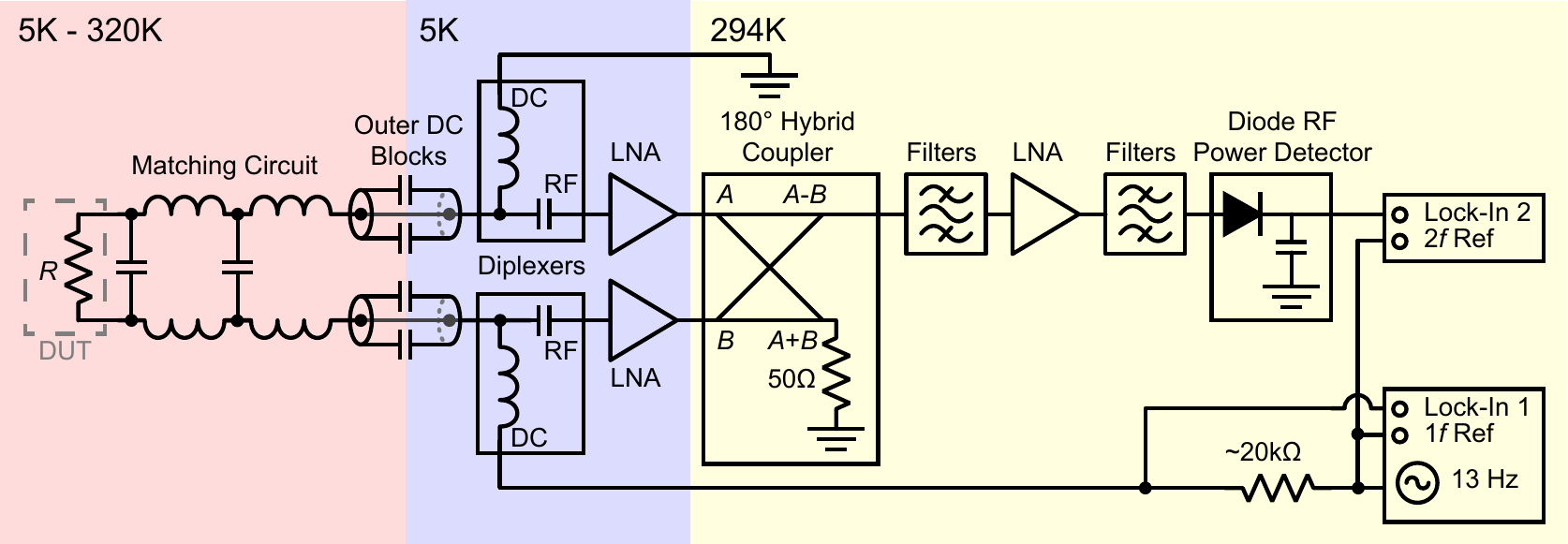}
\caption{\label{fig:full circuit} Schematic diagram of the circuit used for Johnson noise thermometry.}
\end{figure*}

Fig.~\ref{fig:Setup Picture} shows pictures of the thermometer components, and Fig.~\ref{fig:full circuit} shows the corresponding circuit diagram. The thermometer operates on a BLG sample, sitting in a modified variable-temperature cryocooler (Janis SHI-4-5). The sample and MC are on a separate custom-made printed circuit board (PCB) that is attached to the cryostat cold head via an aluminum tube, serving as a weak thermal link, allowing the device to be heated independently of the cold head. We solder the MC using standard chip capacitors and air-core inductors (Piconics MC-series), to avoid temperature-dependence of solid-core inductors. We connect the MC to the sample chip with Al wirebonds and keep the bonding pads small with lateral size of 50~$\mu$m to minimize stray capacitance to the silicon back gate. Coaxial cables connect the rest of the circuit components outside the PCB.

On the cold head, diplexers (Mini-Circuits ZDPLX-2150-S+) separate the low-frequency band ($0-50$~MHz) from the RF band ($> 100$~MHz) for each arm of the circuit. Between the diplexers and hotter sample, we place outer DC blocks (Fairview Microwave SD3462) to improve thermal isolation. We use the low-frequency band for quasi-DC transport, measuring the resistance of and Joule-heating the device with a lock-in technique at a low frequency, typically $\sim17$~Hz. The RF band goes into a cryogenic low noise amplifier (LNA) (Cosmic Microwave Tech CITLF3) with $\sim35$~dB gain. The diplexers and LNAs are held at a constant temperature around 5~K in the cold stage of the cryostat; the LNAs are tuned to have approximately the same gain.

Outside the cryostat at room temperature, a $180\degree$ hybrid coupler (Mini-Circuits ZFSCJ-series) combines the signals from the two individual terminals into a single-ended signal corresponding to the difference of these two signals. Lumped LC Filters (Mini-Circuits SHP-, SLP-, and VLFX- series) form an appropriate passband for the amplified noise, typically around 100-200~MHz. These filters reject frequencies outside the matched band that contain mostly amplifier noise. A room-temperature LNA (Fairview Microwave SLNA-010-050-10-SMA) amplifies the signal by 50~dB to a level above the noise floor of the diode RF power detector (Pasternack PE8000-50). A second set of filters, typically identical to the first set, filters the signal again between the second LNA and the power detector. The power detector takes wide-spectrum RF power and produces a voltage at the output approximately proportional to the input RF power; this voltage is measured both at DC (Agilent 34401A) and at AC at twice the bias frequency, $f_2=2f_1\sim2\times17$~Hz (SR830 lock-in). The nonlinearity of the power detector can be calibrated out with an input of known power from a signal generator.

\subsection{\label{sec:level1}Differential Setup}
Our setup measures noise from the source and drain of the sample in a differential mode. This improves upon the single-ended setup \cite{Crossno2015}, which has one side of the device directly grounded at cold temperature and the other side biased at room temperature, which can induce DC thermoelectric currents. Our differential setup allows independent control of both source and drain contacts at room temperature, removing any thermoelectric effects by the symmetry of the measuring circuit, and allowing for balanced DC measurements or the possibility of DC-floating or DC-grounding any terminals.  Antisymmetric biasing of the source and drain can minimize bias-induced channel gating effect~\cite{XiaomengLiu2017BiasGating} to first order, which would be present in a single-ended setup. The differential setup also rejects common-mode noise picked up along the symmetric arms.

 The differential characteristic impedance of 50~$\Omega$ transmission lines (connected to LNAs) is 100~$\Omega$, which can be viewed as the series-connected input impedances of the two LNAs. This 100~$\Omega$ setup allows easier matches to high-impedance devices than does a single-ended 50~$\Omega$ setup. The $180\degree$ hybrid has an internal 50~$\Omega$ termination for the common mode $A+B$ signal. This termination generates its own Johnson noise that could contribute to the measured noise temperature but is suppressed by the gain of the upstream LNAs~\cite{KC_personal}. It can also be minimized by using a cryogenic hybrid with a single downstream LNA, reducing the number of required LNAs. 

\subsection{\label{Multi-Stage Matching}Multi-Stage Matching}

The goal of the MC is to allow a high efficiency transfer of Johnson noise power from a high-impedance sample to 50~$\Omega$ LNAs, integrated over a wide frequency band. Achieving higher frequency bandwidth has been a standard procedure in electrical engineering with techniques extending far beyond component MCs\cite{pozar2012microwave}. Devices made from gateable 2-dimensional (2D) materials such as BLG, however, require matching circuits that can cover changes in resistance of several orders of magnitude. We thus focus primarily on obtaining the widest resistance range with efficient coupling while maintaining high frequency bandwidth. 

\begin{figure*}
\includegraphics{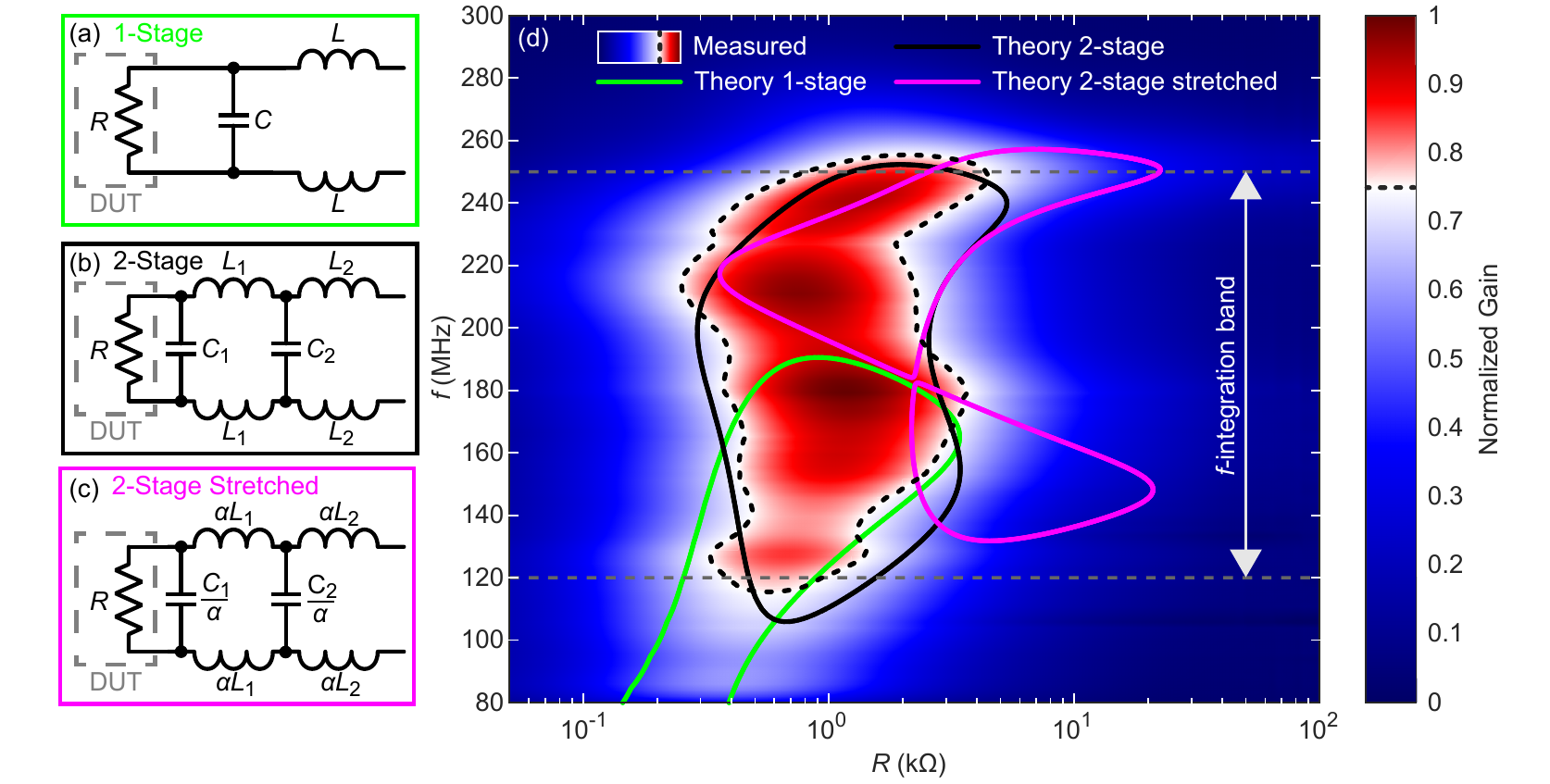}
\caption{\label{fig:MC variants} (a) Standard 1-stage impedance matching. (b) Standard 2-stage impedance matching. (c) Standard 2-stage impedance matching that has been stretched. (d) Measured normalized gain using standard 2-stage matching, as a function of sample resistance and measuring frequency, compared to modeled matching circuits. Colorplot is the measured normalized gain, with white corresponding to 0.75. Drawn contours are corresponding 0.75 coupling efficiency for the measured normalized gain (dotted line), as well as for the three matching circuits in (a)-(c).}
\end{figure*}

The simplest type of MC is a single-stage component matching circuit, shown in Fig.~\ref{fig:MC variants}(a), consisting of two inductors with identical inductance $L$ and a bridging capacitor with capacitance $C$. Here, full power transfer happens at only one point in the resistance-frequency $(R,f)$-plane. At angular frequency $\omega$, the impedance $Z$ of the sample seen by the LNAs through the MC is given by
\begin{align}
Z=2i\omega L+\left( R^{-1}+i\omega C \right)^{-1}.
\label{eq3}
\end{align}

To match at a target angular frequency $\omega_0$ and target sample resistance $R_0$ we equate this impedance to $Z_0=100~\Omega$. This results in MC component values of $C=\sqrt{m-1}/\left( \omega_0 R_0 \right)$ and $2L=\sqrt{m-1}\,Z_0/\omega_0$ , where $m \equiv R_0/Z_0$  is the impedance step ratio.

Using the matched impedance $Z$, the matching efficiency can be expressed as $1-\left| \Gamma \right|^2 =1-\left|\left(Z_0-Z\right)/\left(Z_0+Z\right)\right|^2$. As a function of sample resistance $R$ and frequency $f$, this is given by
\begin{align}
\frac{4m\rho}{\left(1+m\rho \right)^2 +\left[ m-1-\left( 2m^2-3m+1 \right) \rho^2 \right] \Omega^2 +\left( m-1 \right)^2\rho^2\Omega^4}
\label{eq4}
\end{align}
in terms of the normalized parameters $\rho=R/R_0$ and $\Omega=2\pi f/\omega_0$.

A good match is achieved for $R$ and $f$ in a range close to $R_0$ and $f_0$, as shown in Fig.~\ref{fig:MC variants}(d) by the green 75\% matching efficiency contour. The full-width-half-max of the matching surface in $R$ and $f$ is
	
\begin{align}
\Delta R&=4\sqrt{2}R_0 \quad \text{at } \omega=\omega_0
\\
\Delta f& \approx \frac{2}{\sqrt{m}}f_0 \quad \text{at } R=R_0 \quad \text{for } m \gg 1.
\label{eq56}
\end{align}

 These relations show that matching to higher frequency enhances the frequency bandwidth, while matching to higher resistance enhances the resistance range at the cost of decreasing the frequency bandwidth. We define separate $Q$-factors for frequency and resistance using the width of matched regions $\Delta f$ and $\Delta R$, relative to the peak values $f_0$ and $R_0$:
 \begin{align}
 Q_f^{-1}=\frac{\Delta f}{f_0}\approx \frac{2}{\sqrt{m}}; \quad Q_R^{-1}=\frac{\Delta R}{R_0}=4\sqrt{2}.
\label{eq7}
\end{align}

The standard single-stage MC can be further extended to a two-stage MC by adding an additional LC-stage, as shown in Fig.~\ref{fig:MC variants}(b). Compared to a single-stage MC, a two-stage MC allows power transfer over a wider resistance range for a given fixed wide frequency band. The two-stage MC sets the first stage impedance equal to $R_1=\sqrt{R_0 Z_0}$ via the following component values:

\begin{align}
C_1=\sqrt{\sqrt{m}-1}/(\omega_0 R_0), \quad 2L_1=\sqrt{\sqrt{m}-1}\,R_1/\omega_0
\\
C_2=\sqrt{\sqrt{m}-1}/(\omega_0 R_1); \quad 2L_2=\sqrt{\sqrt{m}-1}\,Z_0/\omega_0
\end{align}

The matching efficiency function is given exactly by
\begin{align}
 1-\left| \Gamma \right|^2=\frac{4m\rho}{A+B\Omega^2+C\Omega^4+D\Omega^6+E\Omega^8}
\label{eq8}
\end{align}
where the coefficients in the denominator are given by
\begin{align*}
A&=\left( 1+m\rho \right)^2
\\
B&=- \left( \sqrt{m}-1 \right) \left( 1-2\sqrt{m}-m+ \left( 4m^{3/2}+m-2\sqrt{m}-1 \right) \rho^2 \right)
\\
C&= \left(\sqrt{m}-1 \right)^2 \left(6m\rho^2-\rho^2-1+2\sqrt{m} \left(\rho^2-1\right)\right)
\\
D&=-\left(\sqrt{m}-1\right)^3 \left( \left(4\sqrt{m}+1\right) \rho^2-1\right)
\\
E&=\left( \sqrt{m}-1 \right)^4\rho^2.
\end{align*}

In addition to a perfect-matching solution at $R_0$ and $\omega_0$, we obtain another degenerate perfect-matching solution at $R_0$ and $\omega_1=\omega_0\times\sqrt{\sqrt{m}+1}/\sqrt{\sqrt{m}-1}$.    The 75\% matching efficiency is shown in Fig.~\ref{fig:MC variants}(d) by the black contour line, and full-width-half-maxes and Q-factors are 
\begin{align}
\Delta R&=4\sqrt{2}R_0 \quad\text{ at } \omega=\omega_0,\:\omega_1
\\
\Delta f&\approx \frac{\sqrt{2}}{\sqrt[4]{m}}f_0 \quad\text{ at } R=R_0 \quad\text{ for } m\gg1
\\
Q_R^{-1}&=4\sqrt{2} \quad\text{ at } \omega=\omega_0,\: \omega_1
\\
Q_f^{-1}&\approx \frac{\sqrt{2}}{\sqrt[4]{m}} \quad\text{ at } R=R_0 \quad\text{ for } m\gg1.
\label{eq10}
\end{align}

Like the single-stage case, frequency bandwidth is proportional to the match frequency, resistance range is proportional to the match resistance, and higher match resistance reduces frequency bandwidth. We find that the additional LC stage does not change $Q_R^{-1}$ , but increases $Q_f^{-1}$ for the same $m$, indicating there is significant additional bandwidth. If such a large frequency bandwidth were used with the single-stage MC, the band-integrated resistance range would be much smaller.
	
The frequency bandwidth and resistance range  can be increased further by introducing more stages to the matching circuit. However, additional stages decrease the first-stage capacitance and increase the first-stage inductance values. Mesoscopic devices typically possess stray capacitance due to the presence of heavily doped silicon substrates, sample enclosures and other factors that are difficult to eliminate entirely. With three stages, the first-stage capacitor needs to be smaller than the stray values in our circuits, typically 0.1-0.4~pF. Further stages demand more space for the additional circuitry and the increased inductance, reducing practicality for low temperature setups. For these reasons, we choose to utilize the two-stage circuit. 

\subsection{\label{sec:level1}Stretched Matching Parametrization}
Additional component combinations for 2-stage matching can be obtained by parametrically stretching the standard two-stage matching described in the previous subsection. It is possible to manipulate the perfect matching solution points in the $(R,f)$-plane by scaling the inductors and capacitors via
\begin{align}
L_i\rightarrow \alpha L_i
\\
C_i \rightarrow C_i/\alpha
\end{align}
as shown in Fig.~\ref{fig:MC variants}(c), where $\alpha$ is a parameter with values typically around 2-4, limited by the stray capacitance at the first stage. The resulting effect is demonstrated in Fig.~\ref{fig:MC variants}(d), transforming the black contour into the purple one. Two solutions are shifted to higher resistance $(R\rightarrow \alpha^2 R)$ and spread out in frequency, while one is held fixed at the original match value.  The frequency-band-integrated coupling will thus be enhanced at higher resistance but reduced near the original match. Compared to the standard two-stage match at a correspondingly larger resistance, the stretched match has its solutions spread out more in resistance and frequency and subsequently has a larger frequency-bandwidth, at the cost of reduced coupling near the best-match resistance due to the empty space in the center of the band ($\sim200$~MHz in Fig.~\ref{fig:MC variants}(d)). In addition, the resistance derivatives of the coupling are reduced, reducing parasitic signals (see below).  In principle, the coupling in this setup can be optimized to surpass the standard two-stage match by using advanced filtering strategies, such as a mixer with a tuned local oscillator frequency and tunable low-pass filter, multi-bandpass filtering, or digital signal acquisition for multiband filtering.   

\section{\label{sec:level1}Calibration}
One of the most difficult parts in any noise measurement is performing an accurate calibration due to unknown exact values of amplifier gain, frequency bandwidth, and losses in the system. In our calibration, we combine these unknown prefactors into a total system gain, which we obtain from measuring Johnson noise at several fixed cryostat temperatures set by precalibrated thermometers. The gain is a function of the sample resistance only, as described in the following model.

The noise power going into the power detector in Fig.~\ref{fig:full circuit} is given by
\begin{align}
P_\text{detector}= 
\int 4k_BG_0 \left(\left(1-\left| \Gamma \right|^2 \right)T_\text{samp}+\left| \Gamma \right|^2 T_{\text{N,in}}+T_\text{N,out} \right) df
\label{full_noise_3_chan}
\end{align}
where $G_0=G_0(f)$ is the frequency-dependent effective amplifier gain including the bandpass filters and any losses, $\Gamma=\Gamma(R,f)$ is the reflection coefficient of the sample with respect to the amplifier as defined in Sec.~\ref{sec:Introduction}, $T_\text{samp}$ is the temperature of the sample, and $T_{\text{N,in}}$ and $T_{\text{N,out}}$ are the effective noise temperatures of the noise produced at the input and output terminals of the LNA, respectively. The input-terminal noise of the amplifier reflects off the sample and enters the amplifier, adding to the output noise temperature.

With integration, Eq.~\ref{full_noise_3_chan} can be empirically simplified to
\begin{align}
P_\text{detector}=G\left(T_\text{samp}+T_\text{N} \right)
\label{eq:G_T_TN}
\end{align}
where $G$ and $T_\text{N}$ are the effective band-integrated gain and noise temperatures of the thermometer circuit, which depend only on the sample resistance $R$. The gain $G$ can be obtained by measuring the derivative $dP_\text{detector}/dT_\text{samp}$ by changing the cryostat temperature, as long as the sample resistance stays constant. However, in general the sample resistance also changes with temperature, for which we have developed our calibration technique.
	
\begin{figure}
\includegraphics{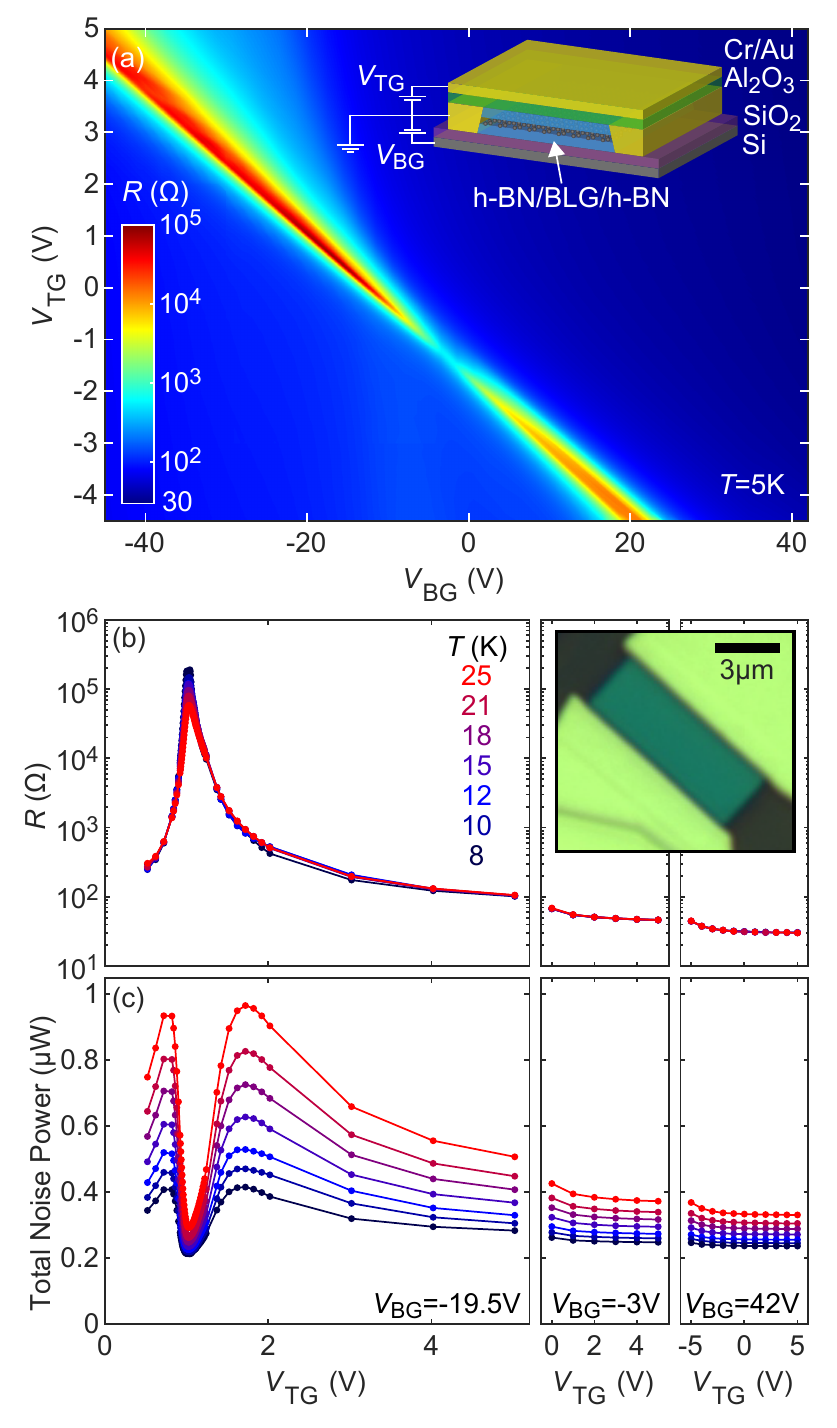}
\caption{\label{fig:Calibration_A}(a) The sample resistance measured at $T=5$~K vs. top gate and back gate voltage ($V_\text{TG},V_\text{BG}$). Inset shows a schematic for the device cross-section. (b, c) Top panel shows resistance as a function of $V_\text{TG}$ at fixed temperatures. Bottom panel shows total noise power measured from the sample as a function of $V_\text{TG}$ at the corresponding temperatures. Each horizontal panel shows data from different fixed back gate voltage. To perform the calibration, we sweep the gates at fixed temperatures to vary the resistance and measure total noise power. Inset: an optical image of the device before top gate deposition.}
\end{figure}
	
The key part of calibration is that the sample resistance $R$ can be varied by gate voltages (or magnetic fields) at several fixed temperatures $T_{samp}$, allowing us to determine $G$ and $T_N$ for any given $R$. In this work, we use BLG as the sample since it can be electrostatically tuned between a metal and a small-gap semiconductor \cite{Castro2007,Oostinga2007,Min2007,McCann2006,Ohta2006,Zhang2009_FengWang_BLGgap,McCann2013Review}. The BLG device was fabricated using standard mechanical exfoliation followed by dry-polymer transfer h-BN encapsulation \cite{Dean2010,Wang2014Thesis,Pizzocchero2016_PPC_hot_pickup} and standard electron beam lithography to build edge-contacting source and drain electrodes \cite{Wang2013}. We use heavily doped silicon substrate as back gate, and a Cr/Au metallic top gate layer deposited after 20~nm atomic layer deposition growth of Al\textsubscript{2}O\textsubscript{3} as gate dielectric. Inset of Fig.~\ref{fig:Calibration_A}(a) and Fig.~\ref{fig:Calibration_A}(c) show a schematic diagram and optical image of the device. Tuning both top gate voltage ($V_\text{TG}$) and back gate voltage ($V_\text{BG}$) to the same sign dopes the BLG channel, while applying gate voltages of opposite sign in the correct proportions opens a gap and places the chemical potential inside the gap. Indeed, Fig.~\ref{fig:Calibration_A}(a) shows the sample resistance $R$ change 30~$\Omega$-100~k$\Omega$ as we tune the chemical potential into the gap at 5~K, creating an ideal testing ground for our variable-resistance JNT.
	
Utilizing the variable resistance, we calibrate our noise thermometer as follows. First, at a fixed cryostat temperature $T_1$, as we vary the top gate voltage $V_\text{TG}$, we simultaneously measure (1) the low-bias differential resistance $R(V_\text{TG})$ of the channel using a lock-in technique and (2) the total noise power $P_\text{detector}(V_\text{TG})$, as shown in Fig.~\ref{fig:Calibration_A}(b) and (c). The goal of this gate voltage sweep is to vary the resistance over as wide a range as possible, so we do it for several back gate voltages to get the full range. This process is repeated for several other temperatures $T_2, T_3,...$.

\begin{figure}
\includegraphics{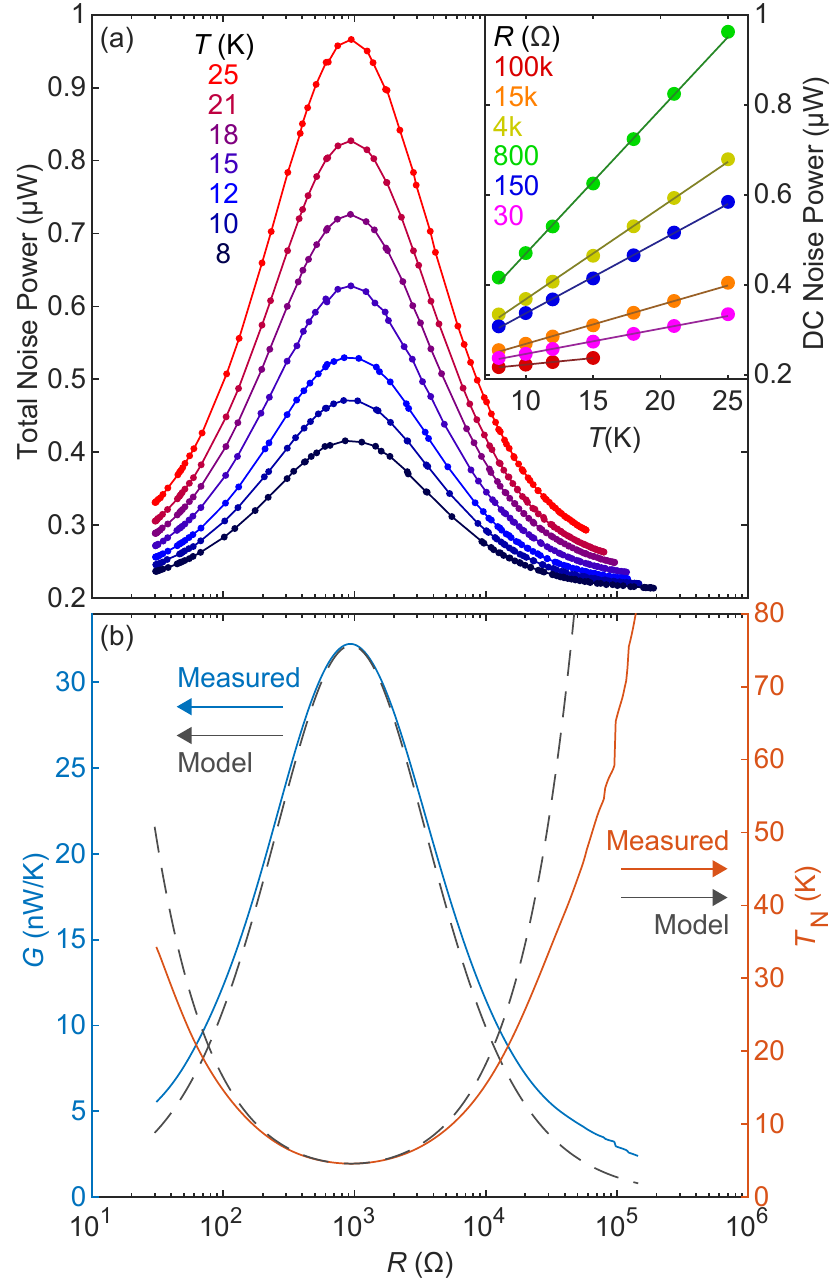}
\caption{\label{fig:Calibration_B} (a) Total noise power plotted vs. sample resistance for the set of measured temperatures. Inset: For any resistance value, we construct a total noise vs. temperature plot with the data measured in (a). The linear fit gives the gain and noise temperature. (b) The measured gain (blue) and noise temperature (red) as a function of sample resistance obtained from data in the inset of (a). The gray dashed curves are the scaled gain and noise temperature for the matching circuit model described in the text..}
\end{figure}

Next, for each temperature $T_i$ the total noise is plotted against the resistance as in Fig.~\ref{fig:Calibration_B}(a). Using interpolation or polynomial fits, we can then obtain a noise power vs. temperature relation for any fixed resistance as shown in Fig.~\ref{fig:Calibration_B}(a) inset. We find the noise power linear in temperature with an offset, consistent with Eq.~\ref{eq:G_T_TN}, for our entire range of measured resistances. For each resistance we then perform a linear fit, obtaining the gain $G$ as the slope and the noise temperature $T_N$ as the horizontal offset. Fig.~\ref{fig:Calibration_B}(b) shows the resulting $G$ and $T_N$ vs. sample resistance $R$. We optimized the MC for $R_0=1$~k$\Omega$, where it shows a peak of $G$ and minimum of $T_N$. Away from this optimized $R$ value, $G$ decreases and $T_N$ increases, reducing our thermometer's precision.

To estimate the relative accuracy, we compare the measured gain value with the calculated $1-\left| \Gamma \right|^2$ value, numerically integrated over the measured frequency band and scaled to equate the peak heights. We use the actual matching circuit component values ($C_1=0.9~\text{pF},\: L_1=240~\text{nH},\: C_2=4.3~\text{pF},\: L_2=76~\text{nH}$) and use the stray capacitance as a fitting parameter to tune the horizontal position of the peak, with $C_\text{stray}=0.12~\text{pF}$. We model the noise temperature with $T_\text{N,out}=4.2$~K and $T_\text{N,in}=1.5$~K as additional fitting parameters. We have plotted this ideal matching as the gray dashed curves in Fig.~\ref{fig:Calibration_B}(b).
	
At the high and low end of the sample resistance ranges, the measured gain deviates from the ideal matching, setting the working resistance range for our thermometer. The deviations can be explained by the presence of an artificial open-circuit gain, estimated at  $\sim$1.6~nW/K by performing a calibration on the matching circuit with no resistive sample connected ($R\rightarrow \infty$). This artifact can be caused by dissipative or lossy circuit elements whose temperature follows that of the sample during calibration, as they will produce their own Johnson noise that varies concurrently with the sample temperature. Additionally, any temperature-dependence of absorption or reflection in variable-temperature circuit elements will change the Johnson noise as the sample is heated. For this reason, we minimize losses between the sample and LNA by using short coaxial cables, and we improve thermal isolation between the sample and other circuit components by using outer DC blocks.

To verify our understanding of the frequency integration in going from Eq.~\ref{full_noise_3_chan} to Eq.~\ref{eq:G_T_TN}, we perform a different frequency-resolved version of the calibration just described. We modify the circuit in Fig.~\ref{fig:full circuit} by removing the filters and replacing the RF power diode with a spectrum analyzer. Instead of measuring the total power $P_\text{detector}$, we now measure the frequency-resolved power spectrum, which is the integrand of Eq.~\ref{full_noise_3_chan} without the filters. We then perform the previously-described calibration method for every $\sim2$~MHz frequency bin. The resulting quantity is $G_0(1-\left|\Gamma\right|^2)$ as a function of $(R,f)$, which we have normalized to unity and plotted in Fig.~\ref{fig:MC variants}(d) as the
colormap with the dotted 0.75 contour. This contour approximately matches the solid black contour from the model for the ideal 2-stage matching circuit; the differences can be explained by a frequency-dependent LNA gain and circuit components whose values do not exactly match the idealized component values.

\section{\label{sec:level1}Electronic Thermal Conductance Measurement Using JNT}
Johnson noise thermometry directly probes the electron temperature of the sample. In graphene at low enough temperature, the electron temperature can de-couple from the lattice temperature, allowing the electrons to be heated to a temperature appreciably higher than that of the lattice \cite{Yigen2013,Yigen2013_2,Fong2013,Crossno2015,Crossno2016}. If electron-electron scattering is strong enough to allow a well-defined electron temperature, electrons in mesoscopic graphene samples obey diffusive quasi-equilibrium transport. In this regime, cooling occurs by electronic diffusion to the bulk metal contacts, dominating cooling by direct heat exchange to phonons. These conditions allow us to use JNT to measure the electronic thermal conductivity ($\kappa$) of monolayer graphene\cite{Fong2012,Fong2013,Crossno2015,Crossno2016}.

In this section, we use our variable-resistance noise thermometry technique on the BLG sample to demonstrate $\kappa$ measurements on a mesoscopic scale. We follow the same assumptions as for monolayer graphene, experimentally confirming that direct phonon cooling becomes appreciable only at higher temperatures.  We measure $\kappa$ as follows: we heat the electrons in the sample with a known amount of power via an AC current, typically at $1f\sim17~\text{Hz}$.  We measure the AC resistance using a lock-in amplifier at $1f$ and noise power modulation at the output of the power detector with a lock-in amplifier at $2f$. Using the calibrated value of the gain at the measured sample resistance, we convert the noise modulation amplitude to a temperature modulation. The ratio of applied Joule power to measured temperature rise then yields the generalized thermal conductance of the sample:
\begin{align}
    G_\text{th,gen}=\frac{P_\text{Joule}}{\Delta T}.
\end{align}

The effective thermal conductance is $G_\text{th}=G_\text{th,gen}/12$ with the factor of 12 arising from the spatial temperature distribution in channel self-heating\cite{Crossno2015,Yigen2013,Yigen2013_2,Fong2013}. The thermal conductivity $\kappa$ is then related by a geometrical aspect ratio.   Measurement with this $2f$ modulation technique is advantageous over applying DC bias and measuring total noise due to reduced susceptibility to drifts of the gain and amplifier noise. 

In addition to the $2f$ component of the noise power signal due to the temperature modulation discussed above, parasitic effects can also arise in the $2f$ signal. In particular, if the differential resistance of the sample changes strongly with the bias voltage $V$ (either due to inherent nonlinearity in the $I$-$V$ curve or due to resistance change caused by heating), then the matching conditions also change with bias, creating modulations of the gain and noise temperature, thus modulating the total noise power output at $2f$. Here, the total $2f$ component of the noise power can then be expressed by the second derivatives:
\begin{align}
    N_{2f}=\frac{1}{4} \left[ \frac{d^2G}{dV^2} \left(T_\text{samp}-T_\text{N,in} \right)+G \frac{d^2T_\text{samp}}{dV^2}\right]_{V=0} V_{1f}^2 +\bigO\left(V_{1f}^4\right)
\end{align}
where $V_{1f}$ is the amplitude of the $1f$ quasi-DC voltage bias, $N_{2f}$ is the amplitude of the $2f$ modulation of the noise, $G$ is the gain as in Eq.~(\ref{eq:G_T_TN}), and $T_\text{N,in}$ is as in Eq.~(\ref{full_noise_3_chan}). Here, $d^2T_\text{samp}/dV^2$ is the term inversely proportional to thermal conductance. Thus, in order to make the quasi-DC thermal transport measurement work, the first term in the bracket must be made small compared to the second term. This goal can be achieved in principle by working in a regime of high gain and small derivatives of gain with respect to the differential resistance, which can be accomplished with careful design of the matching network.   

\section{\label{sec:level1}Results and Discussion}
We perform AC-bias JNT on the device shown in Fig.~\ref{fig:Calibration_A}(a) and measure $G_\text{th}$ as described above. To achieve a large resistance range in a single gate-sweep, we hold the top gate voltage fixed at 1.0~V and sweep the silicon back gate voltage. The bath temperature is fixed at 10~K, and at each back gate voltage value a feedback loop sets an AC bias ($1f$) to maintain a temperature modulation ($2f$) of $200\pm$5~mK rms. 

\begin{figure}
\includegraphics{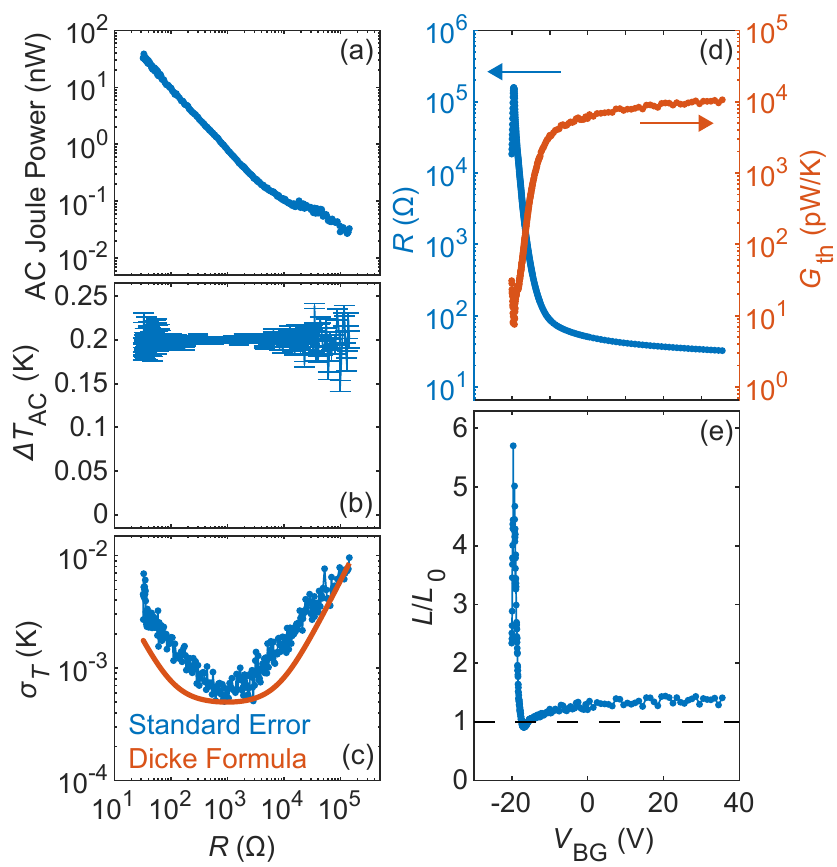}
\caption{\label{fig:DeviceData} (a) The AC bias power applied to the sample as a function of resistance, controlled via a feedback loop to keep the temperature modulation at 200~mK. (b) The measured temperature modulation. The error bars are estimated from the standard deviation of the data before averaging. (c) The uncertainty of measured temperature estimated in (b) is plotted as a function of sample resistance. It approximately follows the prediction from the Dicke radiometer formula (Eq.~(\ref{Dicke_Formula_with_F}) in the text), using 130~MHz bandwidth. The variation as a function of resistance is due to increasing effective amplifier noise temperature away from the optimally matched resistance. (d) The measured resistance and thermal conductance as a function of back gate-voltage, with top gate voltage at 1.0~V. (e) Lorenz ratio computed from the ratio of thermal conductivity to temperature and electrical conductivity. Black dashed line is the Wiedemann-Franz law.}
\end{figure}

Fig.~\ref{fig:DeviceData}(a) and (b) show the AC Joule power and measured temperature modulation as a function of sample resistance. We have computed statistical uncertainties in the measured temperature modulation and plotted those in Fig.~\ref{fig:DeviceData}(b) as error bars and in Fig.~\ref{fig:DeviceData}(c) explicitly. As a reference, we overlaid expected uncertainties from a balanced Dicke Radiometer. From Eq.~\ref{eq1}, the uncertainty is given by\cite{Wait1967}
\begin{align}
    \sigma_T=\frac{F \left( T_\text{samp}+T_\text{N} \right)}{\sqrt{\tau \Delta f_\text{c}}}
    \label{Dicke_Formula_with_F}
\end{align}
with $F=2\sqrt{2}$ for our sine-modulated and sine-correlated radiometer. 

The filters we select in our amplifier chain define a bandwidth that overlaps maximally with the high-coupling region of the matching circuit. For this device, this provides a bandwidth of approximately 130~MHz, from 120 to 250~MHz, as shown in Fig.~\ref{fig:MC variants}(a) by the gray dashed lines. However, the correlation bandwidth $\Delta f_\text{c}$ is slightly smaller and given by \cite{White1996,Wait1967,White1989,Roberts1985,Kittel1977}
\begin{align}
    \Delta f_\text{c}=\frac{\left[ \int_0^\infty \left| G(f) \right| df \right]^2}{\int_0^\infty \left| G(f)\right|^2 df}
\end{align}
where $G(f)$ is the frequency-dependent gain. Our total noise signal is the sum of the three channels shown in Eq.~(\ref{full_noise_3_chan}); each channel has its own uncertainty from its own correlation bandwidth, and the uncertainties add in quadrature to obtain the total uncertainty.

For the entire resistance range measured, our radiometer shows standard errors no larger than about 3 times the predicted uncertainty. The excess standard error can arise from sources such as amplifier gain/noise drifts and additional noise at the input of the lock-in. As the resistance deviates from the 1~k$\Omega$ optimized match, the measurement uncertainty increases due to the increasing effective noise temperature. 

Our 130~MHz noise bandwidth is orders of magnitude larger than bandwidths used in low-frequency  noise thermometry and allows significantly faster measurements. This is an improvement upon single-stage matching, which achieved a 20MHz bandwidth centered at 100MHz for measuring a monolayer graphene device \cite{Crossno2016}. We achieve 650~$\mu$K uncertainty on a 10~K background, equivalent to 65~ppm, with 30~s of averaging.

In Fig.~\ref{fig:DeviceData} we plot in parallel the electrical resistance measured via a standard lock-in technique and the thermal conductance. As the electrical resistance exhibits a strong peak due to opening of the gap by applied gate voltage, the thermal conductance correspondingly falls very sharply. By taking the ratio of the thermal conductance to the electrical conductance ($=R^{-1}$) and the temperature $T$, we obtain the Lorenz number $L=G_{th} R/T$. In a degenerate Fermi liquid, $L$ should be close to the Sommerfeld  value $L_0=\pi^2/3\left(k_B/e\right)^2$. The Lorenz ratio $L/L_0$ then quantifies the degree to which the Wiedemann-Franz law is satisfied in our sample. In our high-resistance gapped BLG, the Lorenz ratio increases up to a maximum measured value of around 5-6. As we move the chemical potential into the valence band, Fermi liquid behavior returns and the Lorenz ratio tends towards 1. Upon further gating, the experimentally measured Lorenz ratio is $\sim$1.3, slightly larger than the expected unity.
	
\section{\label{sec:level1}Conclusion}
We develop JNT applicable to variable resistance samples using high-bandwidth differential noise thermometry. We use a differential two-stage matching circuit to couple a high-resistance device to 100~$\Omega$, the differential characteristic impedance of two coaxial transmission lines. We also describe a modified two-stage matching technique that stretches the matching in the $(R,f)$-plane to allow for variable-resistance operation with larger bandwidth at a reduced coupling at the best-matched resistance. The two-stage matching circuit allows us to obtain 130~MHz of bandwidth centered at 185~MHz, while maintaining high accuracy over two orders magnitude of sample resistance, with reduced accuracy up to four orders of magnitude of resistance. At optimum matching resistance we achieve a temperature measurement uncertainty of 65~ppm in 30~s of averaging time. The uncertainty in the temperature measurement increases as the sample resistance deviates from the optimum value, with the general shape of the curve approximately following the Dicke radiometer uncertainty. Using our noise thermometry, we measured the electronic thermal conductivity of a bilayer graphene sample at 10K by Joule-heating the electrons and measuring the corresponding temperature rise. Our high-bandwidth, variable-resistance differential noise thermometry technique demonstrated in this work can also be applicable to various mesoscopic measurement developments. Fast thermometry can be used for AC calorimetry to measure specific heat of mesoscopic sized samples. Performing noise measurement at finite DC bias voltage will allow us to extend our technique for fast shot noise measurement in future work. 

\section{\label{sec:level2}Acknowledgments}
We thank K.C. Fong and J. Crossno for discussion. This work is supported by ARO (W911NF-17-1-0574) for developing RF technology and characterization, and ONR (N00014-16-1-2921) for device fabrication and measurements. A.T. acknowledges support from the DoD through the NDSEG Fellowship Program. J.W. and P.K. acknowledge support from NSF (DMR-1922172) for data analysis. K.W. and T.T. acknowledge support from the Elemental Strategy Initiative conducted by the MEXT, Japan, Grant Number JPMXP0112101001,  JSPS KAKENHI Grant Number JP20H00354 and the CREST(JPMJCR15F3), JST. The data that support the findings of this study are available from the corresponding author upon reasonable request.


%
%

%



\bibliography{your-bib-file}

\end{document}